\documentclass[sigconf,10pt]{acmart}
\usepackage{xspace}
\usepackage{siunitx}
\usepackage{tabularx}   
\usepackage{booktabs}   
\usepackage{comment}
\usepackage{soul}
\usepackage{makecell}
\usepackage{graphicx}
\usepackage{algorithm}
\usepackage{algpseudocode}
\usepackage{placeins}
\usepackage{xcolor}
\usepackage{wrapfig}

\newcommand{\cname}{\textit{LightFARM}\xspace}

\newcommand{\ppfdunit}{\si{\micro mol\per\meter\squared\per\second}}

\AtBeginDocument{%
  }

\renewcommand\footnotetextcopyrightpermission[1]{} 

\acmSubmissionID{1137}

\begin{document}

\title{\cname: Model Predictive Lighting Control with Battery-Free IoT for Energy-Efficient Indoor Farming}

\author{Hao Yu}
\email{hao.yu4@unsw.edu.au}
\affiliation{%
  \institution{University of New South Wales}
  \city{Sydney}
  \state{NSW}
  \country{Australia}
}

\author{yanxiang wang}
\email{yanxiang.wang@unsw.edu.au}
\affiliation{%
  \institution{University of New South Wales}
  \city{Sydney}
  \state{NSW}
  \country{Australia}
}

\author{Mark Cardamis}
\email{m.cardamis@unsw.edu.au}
\affiliation{%
  \institution{University of New South Wales}
  \city{Sydney}
  \state{NSW}
  \country{Australia}
}

\author{Tianlang Zhang}
\email{tianlang.zhang@unsw.edu.au}
\affiliation{%
  \institution{University of New South Wales}
  \city{Sydney}
  \state{NSW}
  \country{Australia}
}

\author{Yihe Yan}
\email{yihe.yan@unsw.edu.au}
\affiliation{%
  \institution{University of New South Wales}
  \city{Sydney}
  \state{NSW}
  \country{Australia}
}

\author{Hari Ganesan}
\email{hari.ganesan@unsw.edu.au}
\affiliation{%
  \institution{University of New South Wales}
  \city{Sydney}
  \state{NSW}
  \country{Australia}
}

\author{Feiyue Ma}
\email{feiyue.ma@unsw.edu.au}
\affiliation{%
  \institution{University of New South Wales}
  \city{Sydney}
  \state{NSW}
  \country{Australia}
}

\author{Liao Wu}
\email{liao.wu@unsw.edu.au}
\affiliation{%
  \institution{University of New South Wales}
  \city{Sydney}
  \state{NSW}
  \country{Australia}
}

\author{Wen Hu}
\email{wen.hu@unsw.edu.au}
\affiliation{%
  \institution{University of New South Wales}
  \city{Sydney}
  \state{NSW}
  \country{Australia}
}

\renewcommand{\shortauthors}{Yu et al.}

\begin{abstract}
Lighting is the dominant energy load in indoor farming, yet most deployed systems still rely on fixed rule-based or schedule-based control. We present \cname{}, a predictive lighting control framework that couples crop illumination with battery-free sensing for more energy-efficient indoor farming.
\cname{} combines finite-horizon predictive control with compact models of photosynthesis, thermal dynamics, and sensor energy state. The controller adjusts lighting intensity to balance photosynthetic benefit, electrical power consumption, thermal safety, and sensing-energy feasibility. A key design feature is that the same light-emitting diode (LED) fixtures serve both as the photosynthetic light source for crops and as a controllable energy source for self-powered sensor nodes.
We implement \cname{} in a real indoor basil cultivation system and evaluate it through two independent 12-day cultivation trials. Compared with a conventional rule-based baseline, \cname{} reduces lighting energy consumption by approximately 41\% and improves energy productivity from 36.1 to 52.9~$\mathrm{g\,kWh^{-1}}$ and from 41.1 to 60.2~$\mathrm{g\,kWh^{-1}}$ ($\approx 46.5\%$ on average). These results suggest that energy-cooperative predictive lighting control is a promising approach to improving indoor farming efficiency under practical resource constraints, while explicitly accounting for the trade-off between energy savings and crop yield.
\end{abstract}

\maketitle


\section{Introduction}
\label{sec:introduction}

Feeding a projected global population of 10~billion by 2050~\cite{un2024world} is increasingly challenging under climate instability, extreme weather events~\cite{konisky2016extreme}, and declining arable land~\cite{zabel2014global}. 
\emph{Indoor farming}, or controlled-environment agriculture (CEA), has emerged as a promising approach because it enables year-round crop production under tightly regulated temperature, humidity, illumination, and CO$_2$ conditions~\cite{engler2021review}. 
Such control improves production stability and resource efficiency, but also introduces a substantial energy burden.

\begin{figure}[t]
  \centering
  \includegraphics[width=\columnwidth]{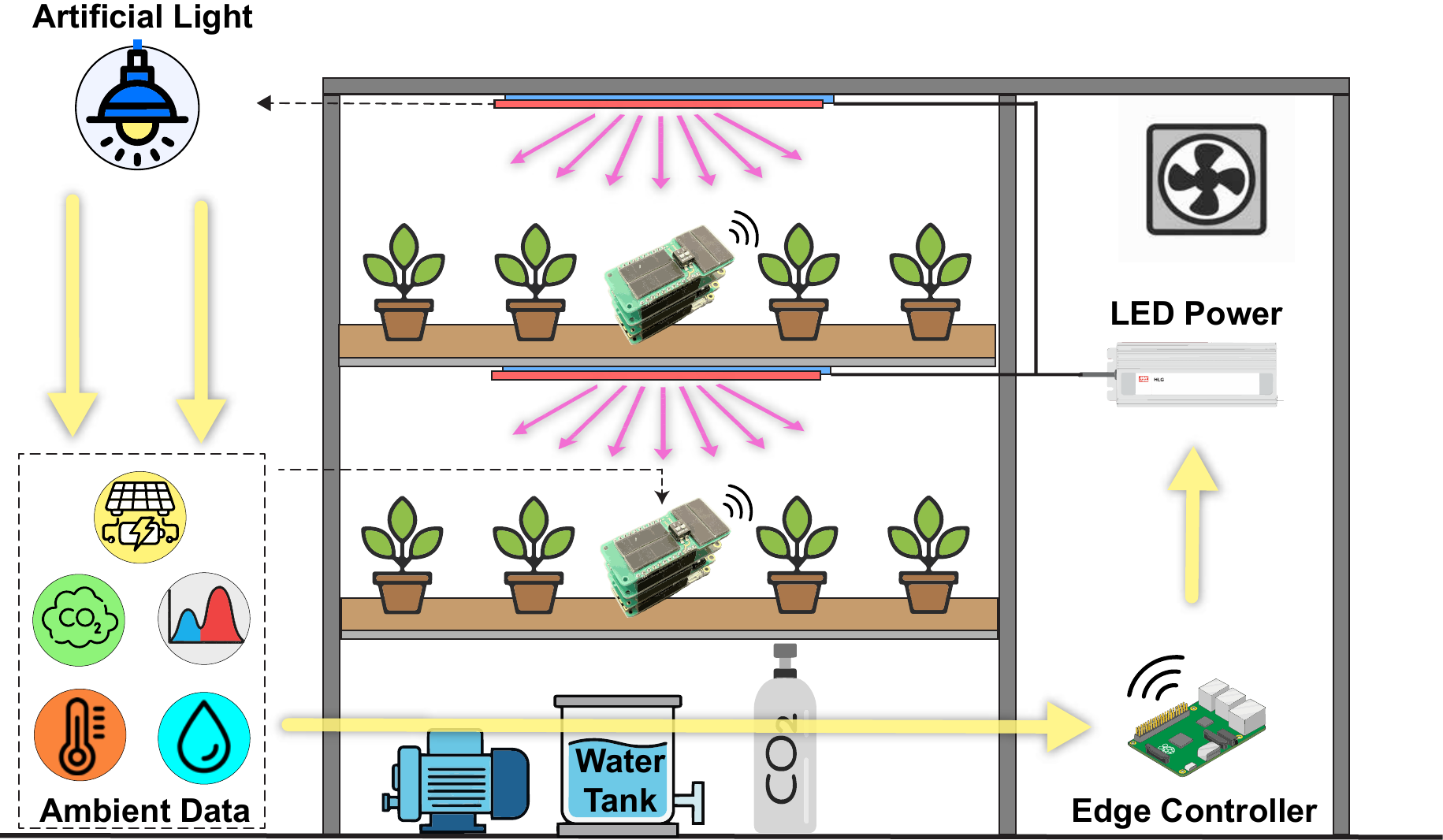} 
  \caption{The conceptual framework of \cname{}: An energy-cooperative predictive lighting control system that jointly optimizes crop growth (photosynthesis) and sensor energy sustainability (energy harvesting) using shared LED infrastructure.}
  \Description{Conceptual overview of LightFARM showing shared LED lighting for crop growth and energy harvesting, battery-free sensing nodes, and a predictive control loop.}
  \label{fig:framework_teaser}
\end{figure}

Indoor farms are commonly divided into \emph{greenhouses}, which supplement sunlight with artificial lighting, and \emph{vertical farms}, which rely primarily on light-emitting diodes (LEDs) as the photosynthetic light source~\cite{stein2021transformative}. 
As illustrated in Figure~\ref{fig:framework_teaser}, a representative vertical farming system integrates LED fixtures, environmental sensors, and control actuators. 
In this work, we propose a co-design approach in which LEDs are treated as a shared resource: they provide the photosynthetic photon flux density (PPFD) required for plant growth while simultaneously serving as a controllable energy source for battery-free Internet of Things (IoT) sensors. 
While this architecture enables fine-grained regulation, its energy footprint remains a major barrier to scalability~\cite{kaiser2024vertical}.

Lighting is a primary control lever in indoor farming, but it is often managed using fixed or schedule-driven policies~\cite{chen2023multi}. 
Such strategies are inefficient because they ignore the nonlinear biological response of plants. 

\begin{figure}[t]
  \centering
  \includegraphics[width=1.0\columnwidth]{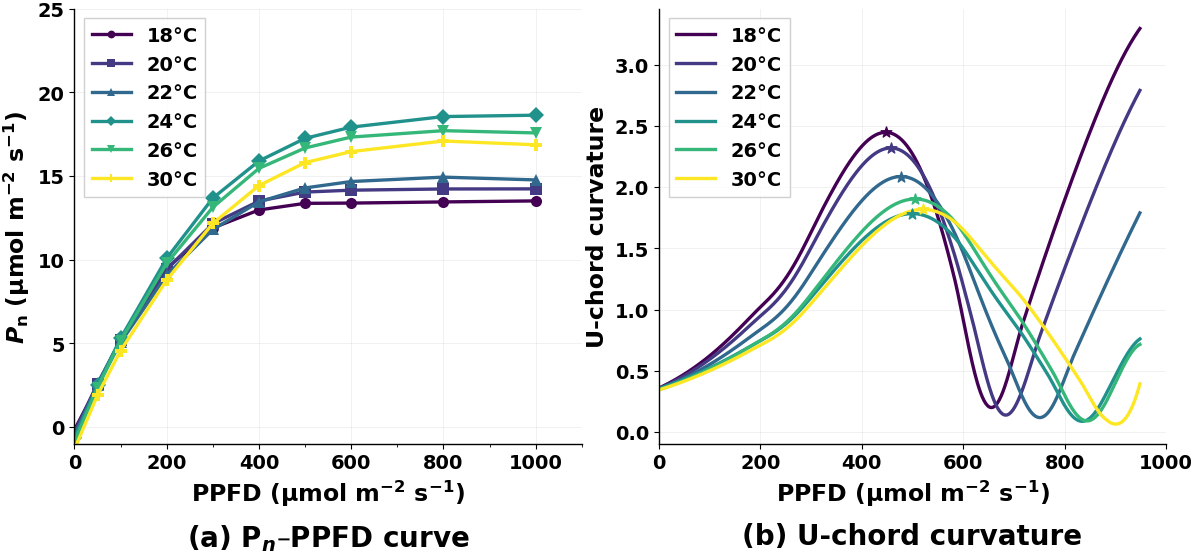}
  \caption{Rationale for lighting optimization: (a) Measured net photosynthetic rate ($P_\mathrm{n}$) shows diminishing returns as light intensity (PPFD) increases; (b) Curvature analysis identifies the ``Light Saturation Point,'' beyond which additional energy input leads to minimal growth gains and excessive heat.}
  \Description{Two plots showing the measured net photosynthetic response to PPFD and the corresponding curvature-based light saturation point used to motivate adaptive lighting control.}
  \label{fig:pn_ppfd_rationale}
\end{figure}

As illustrated in Figure~\ref{fig:pn_ppfd_rationale}(a), the measured net photosynthetic rate ($P_\mathrm{n}$) does not increase linearly with light intensity. 
Once the ``Light Saturation Point'' is reached, additional PPFD provides only marginal physiological benefit while electrical power consumption continues to rise. 
By analyzing the curvature of these response profiles in Figure~\ref{fig:pn_ppfd_rationale}(b), we can identify an efficient operating region for lighting control. 
Beyond this point, excess illumination contributes little to crop productivity but is largely dissipated as heat, thereby increasing the cooling load on heating, ventilation, and air conditioning (HVAC) systems. 
This biological nonlinearity therefore creates a clear opportunity for optimization: by dynamically tracking these efficient regions, the system can reduce lighting energy consumption with limited impact on crop outcome.

\begin{figure}[t]
  \centering
  \includegraphics[width=0.65\columnwidth]{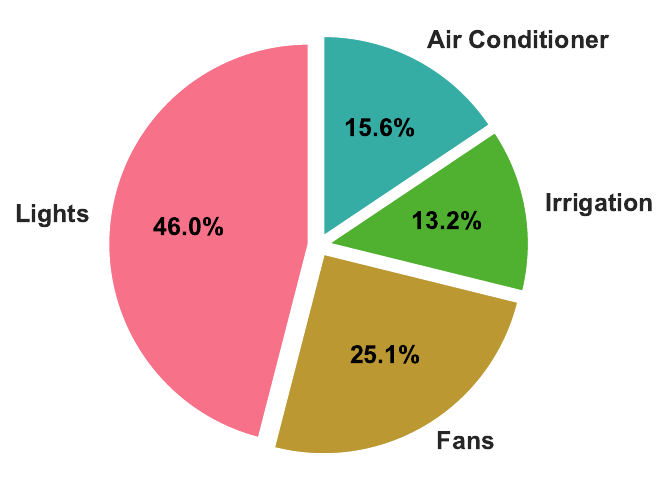}
  \caption{Measured energy consumption of a commercial indoor farm. Lighting accounts for roughly half of total power usage.}
  \Description{Pie chart of commercial indoor-farm energy use showing lighting as the largest share, approximately half of total consumption.}
  \label{fig:energy_pie_chart}
\end{figure}

To quantify the system-level importance of this opportunity, we measured subsystem-level power consumption in a commercial vertical farm. 
As shown in Figure~\ref{fig:energy_pie_chart}, lighting constitutes approximately 50\% of total energy consumption. 
These results identify lighting as the dominant energy consumer and therefore the primary target for system-level optimization.

To address this challenge, we propose \cname{}, a predictive lighting control framework that couples crop illumination with battery-free sensing. 
\cname{} uses Model Predictive Control (MPC) to regulate lighting intensity based on compact models of photosynthesis, thermal dynamics, and sensor energy state. 
The controller balances photosynthetic benefit, electrical power consumption, thermal safety, and sensing-energy feasibility within a unified control loop.

A key aspect of our design is the dual role of LED fixtures.
Unlike battery-free systems that depend on uncontrollable ambient sources (e.g., solar), \cname{} explicitly incorporates the shared lighting--sensing coupling into the control problem.
We develop an analytical energy model that captures optical energy harvesting, storage, and consumption, enabling predictable sensing operation even under dynamic lighting schedules designed for plant growth.
We implement \cname{} as a complete prototype and evaluate it on a perennial basil (\textit{Ocimum gratissimum}) testbed. 
Results show that, relative to a conventional rule-based baseline, \cname{} reduces lighting energy consumption by about 40\% and improves energy productivity by over 46\% across two independent 12-day trials.

In summary, this paper makes the following contributions:
\begin{itemize}
  \item \textbf{Energy-cooperative lighting control formulation:} We formulate indoor farm lighting control as a joint optimization problem that couples crop illumination with battery-free sensor operation under a shared LED infrastructure.
  \item \textbf{Predictive controller with biological and energy models:} We design \cname{}, a finite-horizon predictive controller that jointly considers photosynthetic response, thermal dynamics, and sensing-energy feasibility.
  \item \textbf{Analytical modeling for battery-free sensing:} We develop an analytical model of optical energy harvesting to support predictable operation of self-powered sensor nodes under controllable lighting.
  \item \textbf{Prototype implementation and evaluation:} We demonstrate substantial energy reduction and improved energy productivity in a real-world basil cultivation testbed, providing an explicit energy--yield trade-off analysis.
\end{itemize}

\section{Background Knowledge}
\label{sec:background}

Effective lighting control in indoor farming requires a quantitative understanding of how plants respond to artificial illumination. 
This section summarizes the biological concepts most relevant to \cname{}, focusing on photosynthesis, PPFD, and data-driven modeling of $P_\mathrm{n}$.

\subsection{Crop Growth and Photosynthesis}

Plant productivity in controlled environments is fundamentally governed by \emph{photosynthesis}, the process by which plants convert light energy, CO$_2$, and water into carbohydrates and biomass~\cite{salisbury1988plant}. 
In indoor farming, light functions both as an energy source for carbon fixation and as an environmental signal affecting plant development~\cite{tang2022effect}. 
For lighting control, a key implication is that increasing light intensity does not produce a linear increase in plant productivity. 
Photosynthetic gain typically rises rapidly at low irradiance, then gradually saturates as light increases, while excessive light may contribute little additional productivity but still increase electrical and thermal cost. 
This nonlinear behavior motivates model-based rather than fixed-schedule lighting control in CEA systems.

\begin{figure*}[t]
  \centering
  \includegraphics[width=0.9\textwidth]{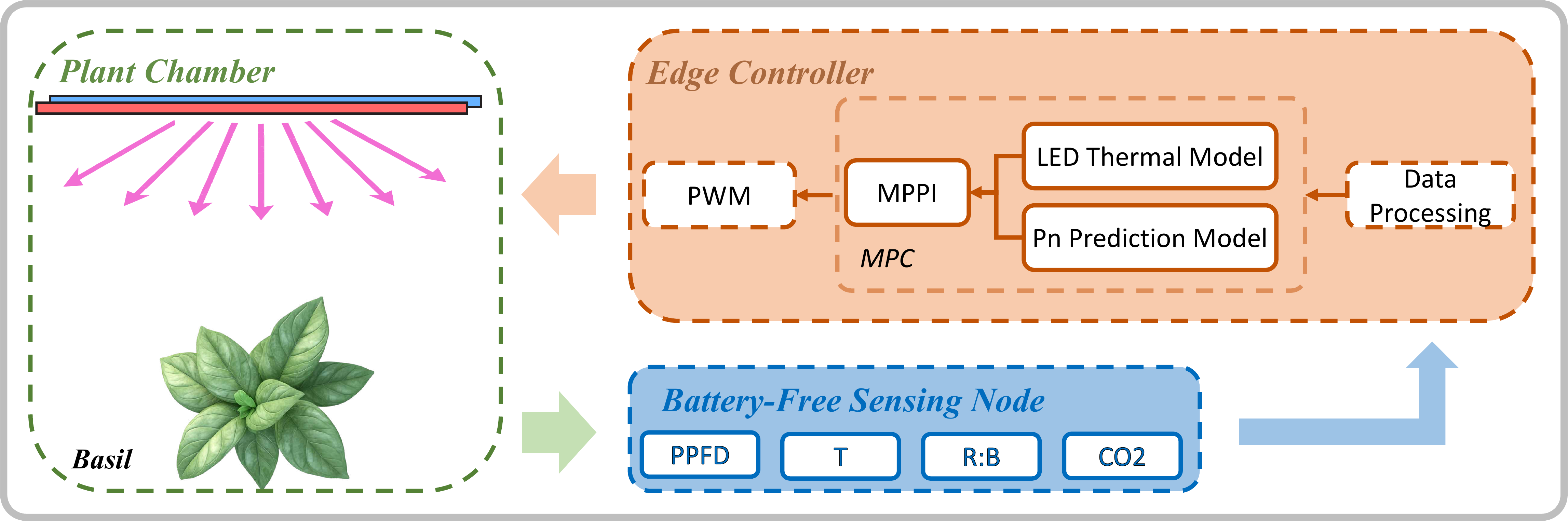}
  \caption{Architecture of \cname, integrating battery-free sensing node, data-driven photosynthesis modeling, and MPC-based lighting control. 
  Environmental data from self-powered nodes feed predictive models of photosynthesis and LED thermal response, which guide illumination scheduling. 
  The same light simultaneously powers both plant growth and sensing, forming an energy-cooperative closed loop.}
  \label{fig:method_system_architecture}
  \Description{System architecture showing feedback loop between battery-free sensing, predictive models, and lighting control.}
\end{figure*}

\subsection{Photosynthetic Photon Flux Density}

PPFD quantifies the instantaneous density of photosynthetically active radiation (PAR) in the 400--700\,nm range~\cite{stallknecht2023designing}. 
Expressed in micromoles of photons per square meter per second (\ppfdunit), PPFD is the standard metric for describing the usable light intensity available to plants. 
Higher PPFD generally increases photosynthetic activity, but the benefit is condition-dependent and often exhibits diminishing returns at moderate-to-high irradiance~\cite{lempiainen2022plants}.

Conventional PPFD measurement relies on laboratory-grade spectroradiometers, which are accurate but costly and impractical for dense embedded deployment. 
To enable distributed low-power monitoring, \cname{} uses low-cost multi-channel spectral sensors such as the AMS AS7341~\cite{ams:as7341}. 
Because such sensors do not directly report PPFD, a calibration model is required to estimate effective PPFD from discrete spectral measurements; this lightweight mapping is introduced later in Section~\ref{sec:ppfdcalib}.

\subsection{Photosynthetic Rate Modeling}
\label{subsec:pn_modeling}

$P_\mathrm{n}$ is the primary metric for carbon assimilation and serves as a proxy for physiological productivity. 
As illustrated by the measured light-response curves in Figure~\ref{fig:pn_ppfd_rationale}(a), the relationship between light, temperature, and growth is highly coupled and nonlinear. 
Beyond the simple diminishing returns of light saturation, we observe a more critical phenomenon: thermal inhibition caused by excessive lighting.

Specifically, increasing PPFD beyond the crop's efficient threshold does not merely result in zero marginal gain; it can actively reduce $P_\mathrm{n}$ by increasing the ambient and leaf temperature.
For example, as shown in Figure~\ref{fig:pn_ppfd_rationale}(a), the $P_\mathrm{n}$ achieved at 26$^{\circ}$C with 500~$\mu$mol~m$^{-2}$~s$^{-1}$ PPFD is notably higher than that achieved at 30$^{\circ}$C with 600~$\mu$mol~m$^{-2}$~s$^{-1}$ PPFD.
This is because the higher electrical power required for 600~PPFD increases the LED heat dissipation ($Q_{\mathrm{heat}}$), which pushes the crop's microclimate into a suboptimal thermal range where metabolic efficiency drops.

This observation directly informs the system design.
In a fully enclosed vertical farm, excess lighting energy is not merely an economic cost; it can also increase thermal stress, raise HVAC demand, and reduce crop performance.
A fixed-schedule policy that pushes high PPFD to maximize growth may, counterintuitively, produce less biomass than a lower-light, temperature-aware strategy.
By capturing these dynamics through our MLP-based $P_\mathrm{n}$ model and incorporating them into the finite-horizon MPC, \cname{} identifies an operating point that balances the trade-off between light-driven growth and heat-induced inhibition.

\section{System Design}
\label{sec:system}


\subsection{System Overview}
\label{subsec:system_overview}

As illustrated in Figure~\ref{fig:method_system_architecture}, \cname{} is a closed-loop indoor-farming system that couples plant lighting control with battery-free sensing.
The central design challenge is that the same LED infrastructure serves two roles simultaneously: providing photosynthetic light for crop growth and supplying controllable energy for self-powered sensor nodes.
Each lighting decision therefore affects plant physiological response, thermal conditions, and sensing availability.

To address this coupling, \cname{} integrates four tightly connected components.
First, a distributed network of battery-free sensing nodes harvests optical energy from the LED lighting and measures local spectral intensity, temperature, humidity, and CO$_2$.
Second, the sensed data feed a plant-side prediction module that estimates $P_\mathrm{n}$ and an LED thermal model that predicts temperature evolution under candidate lighting commands.
Third, a predictive controller computes the pulse-width modulation (PWM) command over a finite horizon to balance plant benefit, electrical energy use, and thermal safety.
Finally, an energy model of the battery-free nodes evaluates whether a candidate lighting schedule can sustain the desired sensing duty cycle, incorporating sensing feasibility into the control loop.

These components form an energy-cooperative feedback loop.
Sensor measurements provide current environmental and energy states.
The predictive models estimate plant and thermal responses under future lighting inputs.
The controller selects a lighting action that determines both crop illumination and the energy harvested by the sensing nodes.
Through this shared-lighting co-design, \cname{} coordinates the biological and cyber-physical subsystems under a unified control framework.


\subsection{Photosynthesis Prediction Module}
\label{sec:photosynthesis_model}

The photosynthesis prediction module provides \cname{} with a real-time estimate of $P_\mathrm{n}$ from low-cost environmental sensing. 
It serves as a virtual plant-state sensor for the controller, replacing direct physiological measurements that are impractical in dense cultivation settings. 
The module consists of two stages: 
(i)~PPFD calibration from multispectral sensor readings, and 
(ii)~prediction of $P_\mathrm{n}$ from PPFD and other environmental variables.

\subsubsection{PPFD Input Calibration}
\label{sec:ppfdcalib}

Because low-cost multispectral sensors do not directly report PPFD, a calibration model is required to estimate effective PPFD from discrete spectral measurements. 
Following prior work on embedded light sensing in controlled environments~\cite{paz2020assessing,mohagheghi2023measuring}, we use a lightweight linear model:
\begin{equation}
  \text{PPFD} = \sum_{i=1}^{n} a_i I_i + b
  \label{eq:ppfd_reg}
\end{equation}
where $I_i$ denotes the intensity measured at spectral channel~$i$, $a_i$ are regression coefficients, $b$ is the intercept, and $n$ is the number of usable spectral channels.

To fit this mapping, we collected paired measurements from an AS7341-based sensing node and a laboratory-grade OHS-310P spectroradiometer~\cite{HopooColor} under programmable red--blue LED lighting. 
The dataset spans multiple PPFD levels and red-to-blue ratios (R:B ratios), enabling the model to capture the range of lighting conditions used in our system. 
The resulting regression model is compact enough for deployment on the microcontroller unit (MCU) and provides accurate real-time PPFD estimation; its quantitative performance is reported later in Section~\ref{subsec:pn_validation}.

\subsubsection{Photosynthetic Rate Prediction}
\label{subsubsec:photosynthesis_prediction_model}

Given the calibrated PPFD, the plant-response estimator predicts $P_\mathrm{n}$ using the feature vector
\begin{equation}
\mathbf{x} = [\text{PPFD},\ \text{R:B},\ T,\ \mathrm{CO}_2]^\top.
\end{equation}
These variables capture the dominant environmental drivers of instantaneous photosynthetic response. 
Before training, all inputs are normalized using min--max scaling:
\begin{equation}
x'_i = \frac{x_i - x_{\min}}{x_{\max} - x_{\min}}.
\end{equation}
Here, $x_{\min}$ and $x_{\max}$ are the minimum and maximum values of feature $x_i$ computed from the training dataset.

To model the nonlinear relationship between $\mathbf{x}$ and $P_\mathrm{n}$, we use a lightweight multi-layer perceptron (MLP), which offers a favorable trade-off between predictive accuracy and inference cost for real-time deployment. 
The model consists of two hidden layers with ReLU activations and a linear output layer:
\begin{equation}
P_\mathrm{n} = f_2\!\left( W_2\, f_1\!\left( W_1 \mathbf{x} + \mathbf{b}_1 \right) + \mathbf{b}_2 \right)
\end{equation}
where $W_1, W_2$ and $\mathbf{b}_1, \mathbf{b}_2$ are trainable parameters, $f_1$ is the ReLU activation, and $f_2$ is linear.

The MLP architecture is selected for its ability to capture coupled effects of light intensity, spectrum, temperature, and CO$_2$ while maintaining low inference latency for real-time deployment.
Model hyperparameters are tuned using particle swarm optimization (PSO) with three-fold cross-validation, and the final model is trained using mean squared error loss. 
Comparisons with alternative regressors and optimizer choices are deferred to Section~\ref{subsec:pn_validation}.

The output $\hat{P}_\mathrm{n}$ is then passed to the predictive controller as a real-time proxy for plant physiological state, allowing lighting decisions to be based on estimated photosynthetic benefit rather than on fixed schedules or light intensity alone.


\subsection{LED Module}
\label{subsec:led_model}

Artificial lighting is a dominant energy consumer in indoor controlled-environment agriculture and directly determines crop photosynthetic performance \cite{pereira2025lighting}. 
Besides providing photosynthetically active radiation, a substantial portion of electrical energy supplied to horticultural lighting systems is dissipated as heat, which contributes to the thermal balance of the cultivation environment \cite{katzin2020greenlight}. 

In our indoor cultivation setup, crops are grown without natural sunlight and no active HVAC system is used for temperature regulation. 
Consequently, the ambient temperature is primarily influenced by the heat generated from LED lighting. 
Under these conditions, the lighting system becomes the dominant controllable thermal input. 
Therefore, \cname{} employs a lightweight control-oriented LED model to predict illumination, electrical power, and temperature evolution under different control actions.

\paragraph{Optical Output.}
Within the operating range of the LED driver, the delivered light intensity is approximately proportional to the PWM duty cycle. 
The resulting PPFD can be expressed as
\begin{equation}
\Phi_{\mathrm{PPFD}}(t)=\Phi_{\max}\,\frac{u(t)}{100},
\label{eq:ppfd_pwm}
\end{equation}
where $\Phi_{\max}$ denotes the maximum PPFD at full duty cycle and $u(t)$ is the PWM command in percentage.

\paragraph{Power and Heat Generation.}
The electrical input power increases proportionally with the PWM duty cycle,
\begin{equation}
P_{\mathrm{in}}(t)=P_{\max}\,\frac{u(t)}{100},
\label{eq:pin}
\end{equation}
where $P_{\max}$ is the rated electrical power of the lighting system. 
Only part of this power is converted into useful radiation, while the remaining energy is dissipated as heat within the lighting system \cite{katzin2020greenlight}. 
This heat term serves as the primary coupling between lighting control and the thermal dynamics of the environment, consistent with prior lumped greenhouse and controlled-environment thermal models~\cite{katzin2020greenlight,katzin2023heating,van2010optimal}.

\paragraph{Thermal Dynamics.}
The temperature response of the cultivation environment is approximated using a lumped first-order thermal model widely used for controlled-environment energy analysis~\cite{katzin2020greenlight,katzin2023heating,van2010optimal},
\begin{equation}
\frac{dT_{\mathrm{amb}}(t)}{dt}
=
\frac{1}{\tau}
\left(
T_{\mathrm{base}} + Q_{\mathrm{heat}}(t)R_{\mathrm{th}} - T_{\mathrm{amb}}(t)
\right),
\label{eq:thermal_ode}
\end{equation}
where $T_{\mathrm{amb}}$ denotes the ambient temperature, $T_{\mathrm{base}}$ represents the baseline temperature without LED heating, $R_{\mathrm{th}}$ is the equivalent thermal resistance, and $\tau$ is the thermal time constant of the air–structure system. 
The effective heating term is modeled as $Q_{\mathrm{heat}}(t)=\eta_{\mathrm{heat}}P_{\mathrm{in}}(t)$, where $\eta_{\mathrm{heat}}$ denotes the fraction of LED electrical power that appears as thermal load in the chamber.
This simplified representation captures the dominant temperature response while remaining computationally efficient for predictive control.

For MPC prediction, the model is implemented in discrete form
\begin{equation}
T_{\mathrm{amb}}(k+1)
=
\left(1-\frac{\Delta t}{\tau}\right)T_{\mathrm{amb}}(k)
+
\frac{\Delta t}{\tau}\left(T_{\mathrm{base}}+Q_{\mathrm{heat}}(k)R_{\mathrm{th}}\right),
\label{eq:thermal_discrete}
\end{equation}
Here, $\Delta t$ is the control interval.
This discrete form allows the controller to anticipate temperature evolution resulting from candidate lighting actions. Since Figure~\ref{fig:pn_ppfd_rationale} suggests that the efficient photosynthetic operating region shifts with temperature, predicting LED-induced thermal dynamics is important not only for safety but also for maintaining operation near the desired plant-response region.


\subsection{Energy Model for the Battery-Free Sensor Node}
\label{subsec:energy_model}

Battery-free sensor nodes operate under intermittent and variable power conditions determined by artificial illumination in indoor farming systems. 
To evaluate whether sensing can be sustained under a candidate lighting action, we develop a control-oriented energy model that captures harvesting, storage, and consumption dynamics. 
The model provides the key interface between LED control and sensing feasibility in \cname{} \cite{kansal2007power,talla2017battery}.

\subsubsection{Harvested Energy Modeling}

Unlike outdoor energy-harvesting systems that rely on broadband solar radiation, the proposed sensing nodes operate under spectrally controlled LED lighting. 
Accordingly, the harvested energy is modeled as a function of spectral photon flux under the applied control action \cite{mohagheghi2023measuring,raghunathan2005design}:
\begin{equation}
e_h(k,u)=
A_{\mathrm{panel}} \, \eta_{\mathrm{cc}} \, \Delta t
\sum_{i=1}^{N_\lambda}
\eta_{\mathrm{pv}}(\lambda_i)\,
\mathrm{PPFD}_{\lambda}(\lambda_i,k,u)\,
\frac{h c N_A}{\lambda_i}\,
\Delta\lambda
\label{eq:eh}
\end{equation}
where $A_{\mathrm{panel}}$ is the photovoltaic area, $\eta_{\mathrm{cc}}$ is the end-to-end conversion efficiency, and $\eta_{\mathrm{pv}}(\lambda)$ is the wavelength-dependent photovoltaic efficiency. 
Here, $\mathrm{PPFD}_{\lambda}(\lambda_i,k,u)$ denotes the spectral PPFD incident on the photovoltaic cell at wavelength bin $\lambda_i$ under lighting action $u$, $N_\lambda$ is the number of wavelength bins, $\Delta\lambda$ is the wavelength-bin width, and $h$, $c$, and $N_A$ are Planck's constant, the speed of light, and Avogadro's constant, respectively.
This formulation makes explicit that the harvested energy depends directly on the lighting action $u(k)$. 
In our platform, the photovoltaic front-end and power-management circuit are abstracted into the effective conversion term $\eta_{\mathrm{cc}}$. 
This approximation is intended for control rather than circuit-level analysis, and is sufficient to capture how changes in LED illumination affect harvestable energy.

\subsubsection{Energy Storage and State Transition}

The harvested energy is buffered in a supercapacitor, whose stored energy is
\begin{equation}
E(k)=\tfrac{1}{2} C_{\mathrm{cap}} V_{\mathrm{cap}}^2(k),
\label{eq:ecap}
\end{equation}
where $C_{\mathrm{cap}}$ is the capacitance and $V_{\mathrm{cap}}(k)$ is the capacitor voltage. 
Following common abstractions in energy-harvesting sensor systems, leakage is neglected relative to active load consumption \cite{kansal2007power}.

To avoid oscillatory switching near the operating threshold, the node state is modeled using hysteresis:
\begin{equation}
\mathrm{State}(k+1)=
\begin{cases}
\mathrm{Active}, & V_{\mathrm{cap}}(k)\ge V_{\mathrm{on}},\\[3pt]
\mathrm{Sleep}, & V_{\mathrm{cap}}(k)\le V_{\mathrm{off}},\\[3pt]
\mathrm{State}(k), & V_{\mathrm{off}}<V_{\mathrm{cap}}(k)<V_{\mathrm{on}}.
\end{cases}
\label{eq:state}
\end{equation}

\subsubsection{Energy Consumption Modeling}

The node energy consumption over interval $k$ is modeled as
\begin{equation}
e_c(k)=
N_{\mathrm{cyc}}(k)\,e_{\mathrm{cyc}}
+
P_{\mathrm{sleep}}\,t_{\mathrm{sleep}}(k),
\label{eq:ec}
\end{equation}
where $N_{\mathrm{cyc}}(k)$ is the number of completed sensing cycles, $e_{\mathrm{cyc}}$ is the energy consumed per sensing cycle, $P_{\mathrm{sleep}}$ is the sleep-mode power, and $t_{\mathrm{sleep}}(k)$ is the accumulated sleep duration. 
Here, $e_{\mathrm{cyc}}$ aggregates sensing, local processing, and wireless transmission energy obtained from platform current traces. 
In practice, one sensing cycle includes sensor sampling, local MCU processing, and one wireless transmission event. 
Aggregating these operations into $e_{\mathrm{cyc}}$ preserves the dominant task-level energy cost while keeping the model compact enough for online control.

\subsubsection{Energy Balance and Feasibility}

Combining harvesting and consumption yields the node energy dynamics
\begin{equation}
E(k+1)=E(k)+e_h(k,u)-e_c(k),
\label{eq:ebalance}
\end{equation}
with the corresponding capacitor-voltage update
\begin{equation}
V_{\mathrm{cap}}(k+1)=\sqrt{\tfrac{2E(k+1)}{C_{\mathrm{cap}}}},
\label{eq:vupdate}
\end{equation}
A candidate lighting action is regarded as feasible for sensing if $E(k+1)\ge E_{\min}$,
where $E_{\min}$ is the minimum stored energy required to sustain the target sensing duty cycle. 
This model therefore determines whether a plant-oriented lighting action can also support battery-free sensing under the current indoor lighting condition.


\subsection{Model Predictive Control}
\label{subsec:mpc}

\cname{} employs an MPC strategy to compute lighting actions that balance plant productivity, electrical energy consumption, and thermal safety over a finite prediction horizon \cite{kouvaritakis2016model}. 
This formulation is well suited to indoor farming because the PWM control input simultaneously affects illumination intensity, predicted photosynthetic response, and ambient temperature. 
Rather than tracking a manually specified photosynthetic target, \cname{} derives an adaptive operating reference from the predicted light-response curve under the current environmental condition. 
As illustrated in Figure~\ref{fig:pn_ppfd_rationale}, the efficient operating region of the photosynthetic light-response curve is not fixed: both the $P_\mathrm{n}$--PPFD response and the corresponding curvature peak vary with temperature. 
This observation motivates an adaptive reference design rather than a constant illumination target.

\paragraph{Adaptive Photosynthetic Reference.}
Let $\hat{P}_\mathrm{n}(\Phi;\theta_k)=f(\Phi,\theta_k)$ denote the predicted $P_\mathrm{n}$ as a function of PPFD $\Phi$ under the current condition $\theta_k$, which includes the local spectral and environmental variables. 
Because photosynthetic light-response curves typically exhibit nonlinear saturation and diminishing marginal gains at high irradiance \cite{marshall1980model,wu2016connecting}, we define the reference PPFD as the point of maximum curvature:
\begin{equation}
\begin{aligned}
\Phi_{\mathrm{ref}}(k)
&=
\arg\max_{\Phi\in[\Phi_{\min},\Phi_{\max}]}
\kappa(\Phi;\theta_k), \\
P_{\mathrm{n},\mathrm{ref}}(k)
&=
\hat{P}_\mathrm{n}(\Phi_{\mathrm{ref}}(k);\theta_k).
\end{aligned}
\label{eq:adaptive_ref}
\end{equation}
Here, $\Phi_{\min}$ and $\Phi_{\max}$ denote the minimum and maximum PPFD values achievable by the LED driver under its PWM bounds.
The function $\kappa(\Phi;\theta_k)$ denotes the curvature of the predicted $P_\mathrm{n}$--PPFD curve. 
The resulting $P_{\mathrm{n},\mathrm{ref}}(k)$ represents a control-oriented operating point at which additional light begins to yield diminishing returns.
Operating near this point helps avoid the deep saturation region, where extra PPFD contributes little additional $P_\mathrm{n}$ and mainly increases electrical consumption and heat generation.

\paragraph{MPPI Objective and Update.}
At each control interval, we evaluate candidate control sequences $\{\mathbf{U}^{(i)}\}_{i=1}^{M}$ over a horizon of length $N$ using the rollout cost
\begin{equation}
\begin{aligned}
J^{(i)}=
\sum_{t=0}^{N-1}
\Big(
& w_P\,[P_{\mathrm{n},\mathrm{ref}}(k+t)-\hat{P}_\mathrm{n}^{(i)}(k+t)]^2 \\
& + w_E\,P_{\mathrm{in}}^{(i)}(k+t) \\
& + w_T\,\phi_T(T_{\mathrm{amb}}^{(i)}(k+t)) \\
& + w_{\Delta u}\,[u^{(i)}(k+t)-u^{(i)}(k+t-1)]^2
\Big),
\end{aligned}
\label{eq:mpc_obj}
\end{equation}
where $w_P$, $w_E$, $w_T$, and $w_{\Delta u}$ weight the photosynthetic tracking, electrical energy, thermal penalty, and control-smoothness terms, respectively. 
Here, $\phi_T(\cdot)$ is a quadratic temperature-violation penalty, and sampled PWM actions are clipped to the actuator bounds during rollout.

We solve this problem using Model Predictive Path Integral (MPPI) control \cite{williams2016aggressive}. 
This choice is motivated by the learned and nonlinear nature of the plant-response model, which makes gradient-based optimization inconvenient, while reinforcement learning would require substantially more interaction data and offers less direct support for operational penalties in our deployment setting.
The first control action is updated by importance weighting:
\begin{equation}
u_{\mathrm{plant}}(0)=
\sum_{i=1}^{M} w^{(i)}u_0^{(i)},
\qquad
w^{(i)}=
\frac{\exp(-J^{(i)}/\lambda)}
{\sum_{j=1}^{M}\exp(-J^{(j)}/\lambda)},
\end{equation}
where $J^{(i)}$ is the rollout cost of the $i$-th candidate sequence and $\lambda$ is the MPPI temperature parameter.
Here, $u_0^{(i)}$ denotes the first PWM action in the sampled candidate sequence $\mathbf{U}^{(i)}$.

\paragraph{Receding-Horizon Execution.}
Only the first optimized control input is applied to the LED driver. 
At the next control interval, sensor feedback is updated, the adaptive photosynthetic reference is recomputed, and the MPPI optimization is repeated. 
This receding-horizon strategy enables online adaptation to changing indoor environmental conditions.

\paragraph{Role in the Full System.}
The MPPI-based MPC forms the plant-centered decision-making core of \cname{}. 
Its output determines the lighting action that best balances photosynthetic benefit, energy consumption, and thermal safety. 
Battery-free sensing feasibility is then checked using the energy model in Section~\ref{subsec:energy_model}, and a minimal correction is applied only when necessary to sustain the target sensing duty cycle.


\subsection{Integrated Plant Growth and IoT Node Energy Co-Design}
\label{subsec:co_design}

The shared LED infrastructure in \cname{} serves two functions simultaneously: it provides the photosynthetic light required for crop growth and the optical energy harvested by the battery-free sensing nodes.
For this reason, lighting control and sensing sustainability cannot be treated independently.
However, in our design, plant productivity remains the primary control objective, while sensing support is enforced as a secondary feasibility requirement.

Accordingly, the co-design mechanism operates as a coordination layer on top of the plant-centered MPPI controller described in Section~\ref{subsec:mpc}. 
At each control interval, the controller first computes a lighting action that balances photosynthetic benefit, electrical energy use, and thermal safety. 
The battery-free energy model is then used to check whether the same action can sustain the desired sensing duty cycle. 
If the predicted next-step stored energy falls below $E_{\min}$, a non-negative PWM compensation term $\Delta u_{\mathrm{sensor}}$ is added to the plant-oriented command. Formally, $\Delta u_{\mathrm{sensor}}$ is the smallest non-negative increment that restores the sensing-energy constraint, i.e., $E(k)+e_h(k,u_{\mathrm{plant}}(0)+\Delta u_{\mathrm{sensor}})-e_c(k)\ge E_{\min}$. This correction enforces sensing feasibility with the smallest possible increase in illumination.

This execution flow is summarized in Algorithm~\ref{alg:co_design}. 
The resulting strategy preserves the main plant-oriented objective of the controller while exploiting the shared-lighting architecture to maintain sustainable battery-free sensing.
$\Delta u_{\mathrm{sensor}}$ (Lines 9 and 10) therefore acts as a feasibility safeguard rather than a primary driver of the lighting schedule. When the harvested-energy margin becomes small, the preferred mitigation is to reduce sensing duty cycle or temporarily tolerate sparser updates, instead of maintaining elevated LED output solely for communication support.

\begin{algorithm}[t]
\caption{Plant-Centered MPC with Sensor-Energy Feasibility Correction}
\label{alg:co_design}
\begin{algorithmic}[1]
\While{system active}
    \State Acquire sensor and energy-state data $(\mathrm{PPFD}, T, \mathrm{CO}_2, E(k))$
    \State Compute the adaptive photosynthetic reference $(\Phi_{\mathrm{ref}}(k), P_{\mathrm{n},\mathrm{ref}}(k))$
    \State Solve the MPPI optimization in Eq.~(\ref{eq:mpc_obj}) to obtain the plant-oriented command $u_{\mathrm{plant}}(0)$
    \State Predict harvested energy $e_h(k,u_{\mathrm{plant}}(0))$ via Eq.~(\ref{eq:eh})
    \State Predict node consumption $e_c(k)$ via Eq.~(\ref{eq:ec})
    \State Compute $E(k+1)=E(k)+e_h(k,u_{\mathrm{plant}}(0))-e_c(k)$
    \If{$E(k+1)<E_{\min}$}
        \State Compute the smallest non-negative $\Delta u_{\mathrm{sensor}}$ such that $E(k)+e_h(k,u_{\mathrm{plant}}(0)+\Delta u_{\mathrm{sensor}})-e_c(k)\ge E_{\min}$
        \State $u_{\mathrm{final}}(0)\gets u_{\mathrm{plant}}(0)+\Delta u_{\mathrm{sensor}}$
    \Else
        \State $u_{\mathrm{final}}(0)\gets u_{\mathrm{plant}}(0)$
    \EndIf
    \State Apply $u_{\mathrm{final}}(0)$ to the LED driver
\EndWhile
\end{algorithmic}
\end{algorithm}

Through this light-centric coordination mechanism, \cname{} maintains battery-free sensing only when necessary and with minimal interference to the primary plant-oriented control objective. 
In this way, the same LED action supports both crop illumination and sensing sustainability, while preserving the intended energy--yield trade-off of the predictive controller.

\section{Evaluation}

\subsection{Experimental Setup and Metrics}
\label{subsec:exp_setup_metrics}

We evaluate \cname{} through a chamber-scale study combining real deployment measurements with model-validation experiments. 
Our evaluation has three objectives: 
(1) to assess the end-to-end performance of \cname{} under realistic indoor farming conditions; 
(2) to quantify its energy--yield trade-off relative to a representative rule-based baseline; and 
(3) to validate the photosynthesis predictor and the practical feasibility of battery-free sensing in the deployed system. 
Perennial basil was selected as the test crop because of its well-characterized photosynthetic response under controlled lighting.
\begin{table}[b]
    \caption{Parameters used in the MPC controller.}
    \centering
    \begin{tabular}{@{}clclcl@{}}
        \toprule
        $N$ & 6 & $R_\text{th}$ & 0.05 & $w_P$ & 50 \\
        $M$ & 700 & $w_E$ & 25 & $w_T$ & 1000 \\
        \bottomrule
    \end{tabular}
    \label{tab:placeholder}
\end{table}
\paragraph{Prototype and Controller Configuration.}
Before deployment, we developed a digital-twin simulator integrating the LED thermal model, photosynthesis predictor, and MPC/MPPI controller for parameter tuning and stability verification. 
The simulator was used to calibrate thermal resistance, tune controller weights, and verify controller convergence under representative indoor conditions. 
The final controller parameters are listed in Table~\ref{tab:placeholder}.

\begin{figure}[t]
  \centering
  \includegraphics[width=1.0\columnwidth]{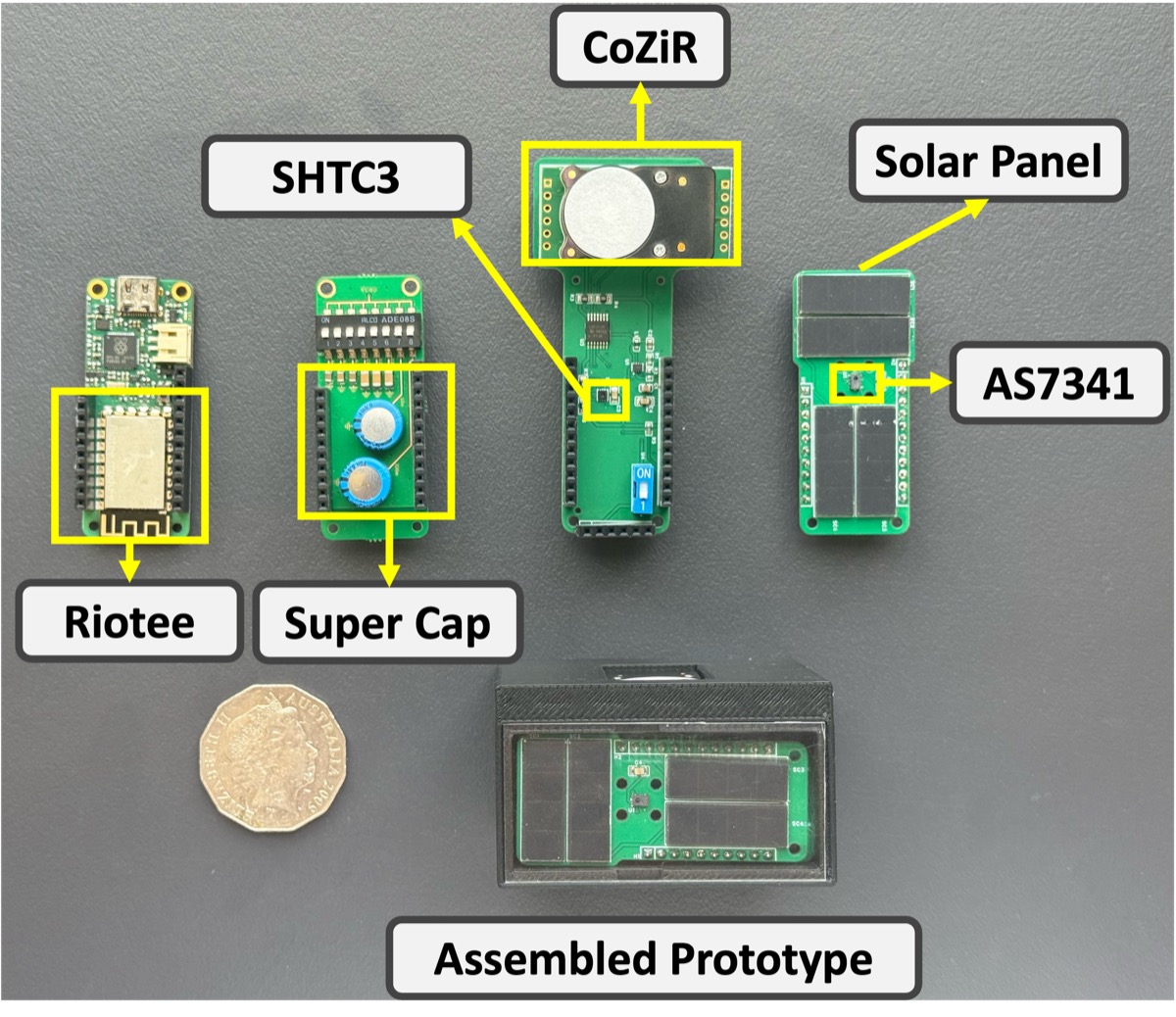}
  \caption{Battery-free sensing node based on the Riotee platform, integrating spectral, CO$_2$, and environmental sensors, with a 3D-printed diffuser enclosure (bottom).}
  \Description{Photograph of the battery-free sensing node hardware and its diffuser enclosure, showing the integrated sensing and energy-harvesting platform.}
  \label{fig:riotee}
\end{figure}

\begin{figure}[t]
  \centering
  \includegraphics[width=1.0\columnwidth]{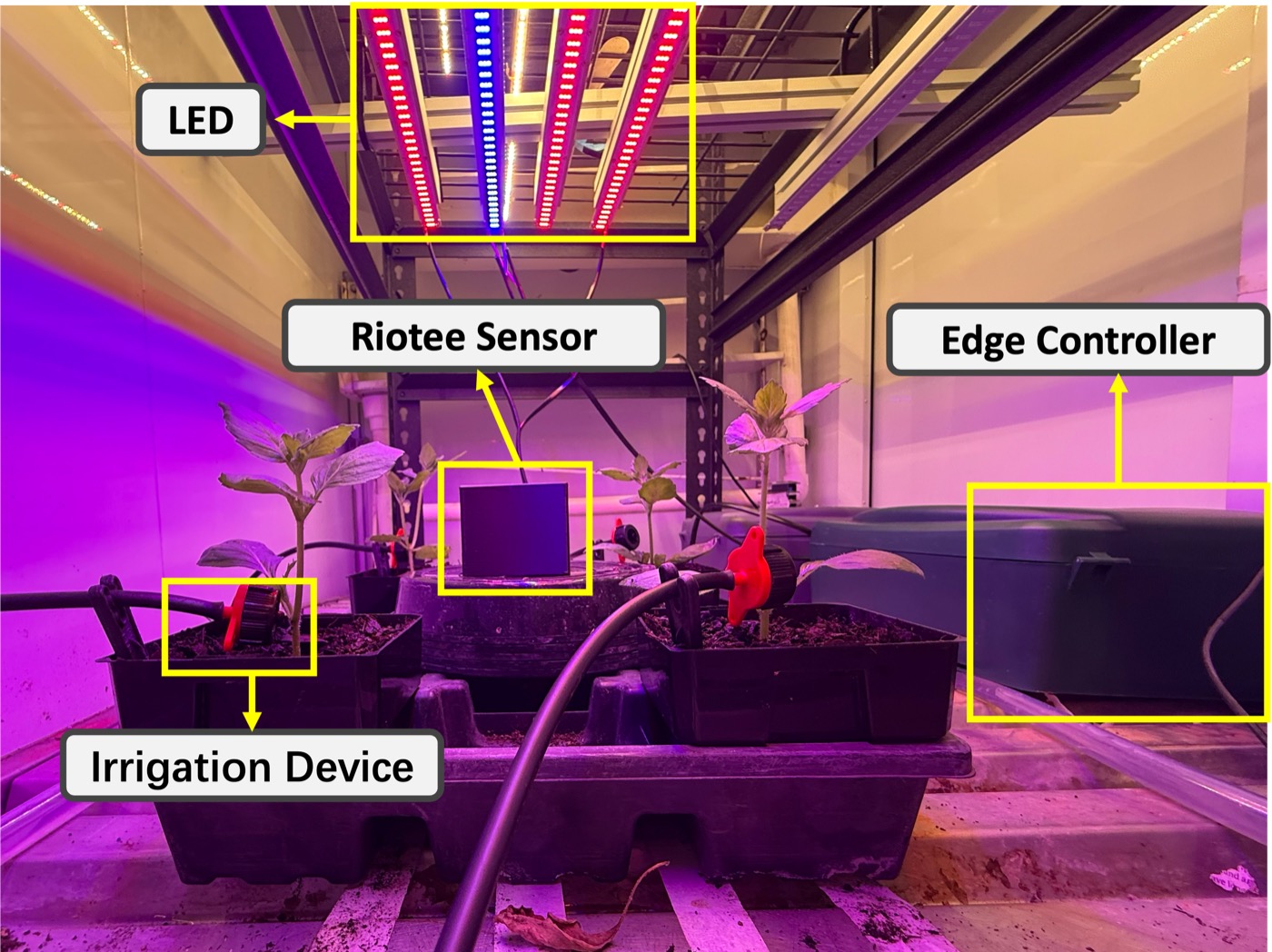}
  \caption{Indoor farming testbed used for evaluation. Each chamber includes four basil plants, custom red--blue LED bars, and a centrally placed battery-free sensing node for environmental monitoring.}
  \Description{Photograph of the indoor farming testbed showing two growth chambers with basil plants, overhead LED bars, and centrally placed sensing nodes.}
  \label{fig:testbed}
\end{figure}

The deployed prototype consists of a battery-free sensing node and an edge controller. 
The sensing node is built on the Riotee platform~\cite{geissdoerfer2024riotee}, harvesting optical energy from the LED illumination through a photovoltaic cell, a MAX20361 power-management IC, and a 100~mF supercapacitor. 
It integrates an AS7341 multispectral sensor, a CozIR-Blink-N CO$_2$ sensor, and an SHTC3 temperature--humidity sensor. 
The edge controller runs on a Raspberry~Pi~4, receives sensor data via Bluetooth Low Energy (BLE), and executes Algorithm~\ref{alg:co_design} in real time. 
Figures~\ref{fig:riotee} and~\ref{fig:testbed} illustrate the sensing node and the chamber deployment.

\paragraph{Experimental Setup.}
The testbed consists of two identical indoor growth chambers (920~$\times$~1500~$\times$~540~mm), each equipped with custom red--blue LED bars (660~nm red, 450~nm blue) and a centrally placed battery-free sensing node. 
The LED bars were mounted 450~mm above the canopy to provide approximately uniform illumination.

The rule-based baseline was configured to deliver approximately 450~\ppfdunit{} at canopy height. 
As shown in Figure~\ref{fig:pn_ppfd_rationale}, the measured $P_\mathrm{n}$--PPFD response exhibits a clear knee region in the moderate PPFD range. 
This observation is also broadly consistent with prior studies reporting diminishing photosynthetic return for basil beyond roughly 400--500~\ppfdunit{}~\cite{pennisi2019unraveling,knight2025photon}. 
We therefore use 450~\ppfdunit{} as a representative fixed baseline rather than a claim of universal optimality.

The two chambers were operated in parallel, one under \cname{} and the other under fixed rule-based lighting. 
Each chamber contained four perennial basil plants. 
Both chambers shared the same irrigation schedule, spectral setting, and surrounding indoor conditions, so that the primary difference was the lighting control policy.

Each experiment lasted 12~days and focused on the vegetative-to-pre-flowering stage, which matches the regime covered by the current $P_\mathrm{n}$ dataset and predictor.
This choice keeps the evaluation consistent with the modeled growth regime while covering a physiologically relevant phase for adaptive lighting control.

\paragraph{Metrics.}
We evaluate \cname{} along four dimensions: lighting energy consumption, crop outcome, photosynthesis-model accuracy, and battery-free sensing feasibility. 
Together, these metrics characterize both the energy--yield trade-off of the lighting controller and the practical feasibility of sensing under harvested energy.

Energy performance is quantified by the average and total lighting energy consumed during each cultivation run, while crop outcome is measured using final fresh biomass. 
To jointly assess yield and lighting cost, we define the energy productivity (EP) as the ratio of final fresh mass (g) to total lighting energy (kWh), i.e., EP = Final Fresh Mass (g) / Total Lighting Energy (kWh), which measures the biomass produced per unit of lighting energy. 
Photosynthesis-model accuracy is evaluated using the coefficient of determination ($R^2$), root mean squared error (RMSE), and mean absolute error (MAE), and sensing feasibility is assessed through power consumption and long-term capacitor-voltage behavior under the deployed indoor lighting schedule.


\subsection{End-to-End System Performance}
\label{subsec:system_performance}

In this subsection, we evaluate the end-to-end performance of \cname{} under real cultivation conditions. 
The experiments compare \cname{} with a representative rule-based baseline under identical environmental settings, including a 16/8~h light/dark photoperiod, a constant R:B ratio of 5:1, and the same irrigation schedule. 
The main experimental difference between the two chambers is therefore the lighting control strategy.

\subsubsection{Overall Energy and Yield Performance}
\label{subsubsec:overall_performance}

To evaluate the consistency of the proposed controller, we conducted two independent 12-day basil experiments under identical operational conditions. 
The results are summarized in Tables~\ref{tab:energy_two_runs} and~\ref{tab:ep_two_runs}. Across both runs, the rule-based baseline consumed \textbf{13.824~kWh}, whereas \cname{} consumed \textbf{8.255~kWh} in Run~1 and \textbf{8.048~kWh} in Run~2, corresponding to lighting-energy reductions of \textbf{40.3\%} and \textbf{41.8\%}, respectively. 
The close agreement between runs indicates that the energy-saving behavior is reproducible.

Fresh biomass was lower under \cname{} than under the rule-based baseline in both runs, indicating that the energy savings were accompanied by a measurable yield trade-off. 
In Run~1, fresh mass decreased from \textbf{498.85~g} to \textbf{437.05~g}, and in Run~2 from \textbf{568.75~g} to \textbf{484.50~g}. 
However, the reduction in lighting energy was proportionally larger, leading to consistent improvements in EP: from \textbf{36.1} to \textbf{52.9~g/kWh} in Run~1 and from \textbf{41.1} to \textbf{60.2~g/kWh} in Run~2, corresponding to nearly identical EP gains of about \textbf{46\%}.

\begin{table}[t]
\centering
\small
\caption{Lighting energy consumption across two 12-day runs.}
\label{tab:energy_two_runs}
\begin{tabular}{lccc}
\toprule
\textbf{Run \& Method} &
\textbf{$P_\text{avg}$ (W)} &
\textbf{Energy (kWh)} &
\textbf{Saving (\%)} \\
\midrule
Run 1 -- Rule-based & 72.00 & 13.824 & --   \\
Run 1 -- \cname{}   & 43.00 & 8.255  & 40.3 \\
Run 2 -- Rule-based & 72.00 & 13.824 & --   \\
Run 2 -- \cname{}   & 41.92 & 8.048  & 41.8 \\
\bottomrule
\end{tabular}
\end{table}

\begin{table}[t]
\centering
\small
\caption{Fresh biomass and EP across two runs.}
\label{tab:ep_two_runs}
\begin{tabular}{lccc}
\toprule
\textbf{Run \& Method} &
\textbf{Mass (g)} &
\textbf{EP (g/kWh)} &
\textbf{$\Delta$EP (\%)} \\
\midrule
Run 1 -- Rule-based & 498.85 & 36.1 & --   \\
Run 1 -- \cname{}   & 437.05 & 52.9 & 46.5 \\
Run 2 -- Rule-based & 568.75 & 41.1 & --   \\
Run 2 -- \cname{}   & 484.50 & 60.2 & 46.4 \\
\bottomrule
\end{tabular}
\end{table}

\subsubsection{Adaptive Control Dynamics}
\label{subsubsec:control_dynamics}

\begin{figure}[t]
  \centering
  \includegraphics[width=1.05\linewidth ]{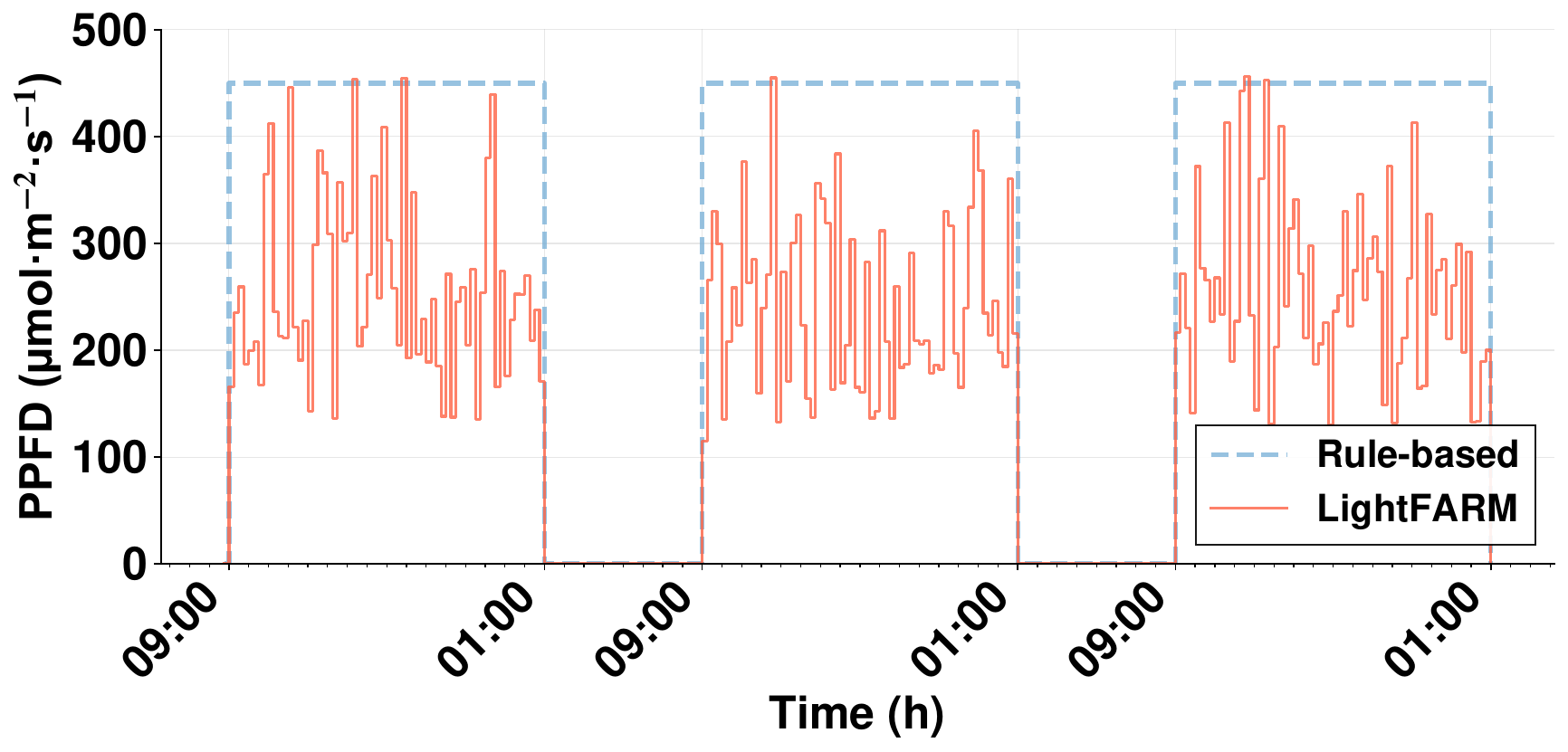}
  \caption{PPFD trajectories over three consecutive days under rule-based and \cname{} control.}
  \Description{Time-series plot comparing PPFD under the rule-based baseline and LightFARM over three consecutive days, with the proposed controller showing a varying profile.}
  \label{fig:ppfd_control}
\end{figure}

Figure~\ref{fig:ppfd_control} compares the PPFD trajectories over a representative three-day window. 
The rule-based strategy maintained a constant illumination of 450~\ppfdunit{}, whereas \cname{} dynamically modulated PPFD according to predicted plant response and thermal feedback. 
The resulting nonuniform yet repeatable profile indicates stable adaptive control behavior.

\begin{figure}[t]
  \centering
  \includegraphics[width=\linewidth]{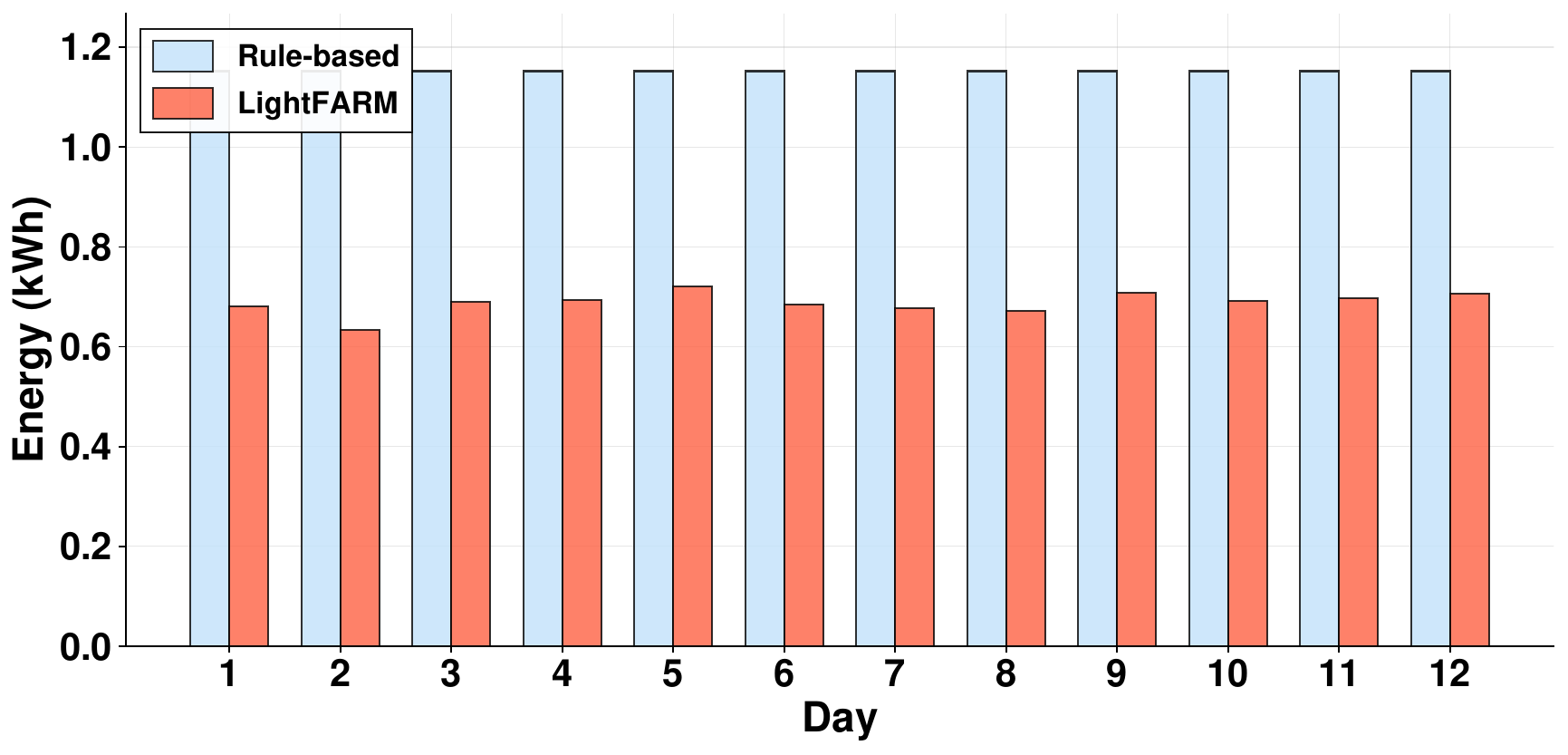}
  \caption{Daily lighting energy consumption comparison between rule-based and \cname{} control.}
  \Description{Time-series plot of daily lighting energy consumption showing consistently lower power under LightFARM than under the rule-based baseline.}
  \label{fig:power_time}
\end{figure}

Figure~\ref{fig:power_time} shows the daily lighting power over the 12-day period for one representative run. 
Compared with the fixed baseline, \cname{} maintained lower average power levels throughout the experiment, demonstrating the benefit of predictive intensity modulation for reducing electrical energy demand.

\begin{figure}[t]
  \centering
  \includegraphics[width=1\linewidth]{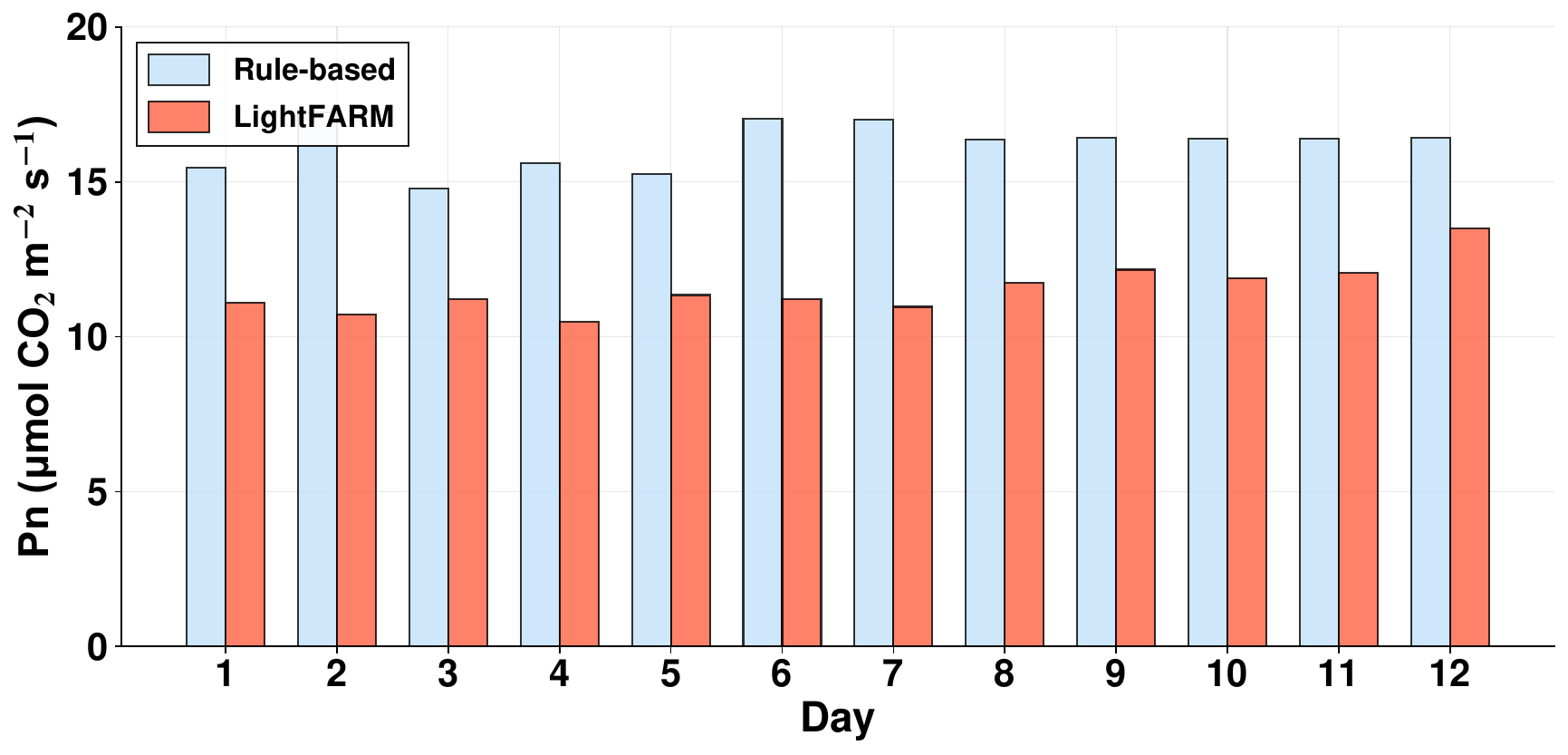}
  \caption{Daily $P_\mathrm{n}$ comparison between \cname{} and rule-based control.}
  \Description{Time-series plot comparing predicted $P_\mathrm{n}$ under LightFARM and the rule-based baseline over the cultivation period.}
  \label{fig:pn_time}
\end{figure}

Figure~\ref{fig:pn_time} shows the temporal evolution of the predicted $P_\mathrm{n}$ for the same run. 
Although \cname{} operated at a lower instantaneous $P_\mathrm{n}$ than the rule-based baseline (\textbf{12.8 vs 16.1 $\mu$mol~CO$_2$~m$^{-2}$~s$^{-1}$}), the response remained stable across photoperiods.

\subsection{Simulation Study}
\label{subsec:simulation}

To further examine the operating characteristics of \cname{}, we conduct a trace-driven simulation study using environmental logs collected from the real deployment. 
The recorded ambient temperature and CO$_2$ traces are replayed offline through the calibrated models, allowing controlled analysis without requiring additional multi-day cultivation experiments.

\paragraph{Baselines and Oracle Advantage.}
In this study, the rule-based and \cname{} results are taken directly from their respective deployment runs.
To isolate the impact of temporal light allocation, we construct a Mean-PPFD policy as an iso-energetic baseline. 
This baseline delivers a constant illumination level set to the exact average PPFD that \cname{} consumed over the entire deployment period. 
It is important to note that the Mean-PPFD policy serves as an offline oracle with an unrealistically favorable assumption in this comparison.
In a real-world scenario, such a perfect mean cannot be determined in advance, as it requires knowledge of the total energy budget and future environmental conditions.

\paragraph{Results Analysis.}
Table~\ref{tab:lighting_sim} summarizes the comparison results. 
For clarity, the relative gap in the last column is computed using integrated photosynthetic productivity per unit lighting energy, i.e., $P_{n,\mathrm{int}}/E$, with \cname{} used as the reference. Compared with the rule-based controller, \cname{} reduces lighting energy from 13.824~kWh to 8.255~kWh while the rule-based baseline remains 30.2\% lower on this metric. 
Crucially, even when compared against the Oracle-aided Mean-PPFD baseline under the same energy budget, the Mean-PPFD policy remains 11.0\% lower than \cname{} in integrated photosynthetic productivity per unit lighting energy. 
These results highlight the advantage of model-predictive control: by dynamically tracking efficient operating regions as temperature and CO$_2$ vary, \cname{} uses the same photon budget more effectively than a constant-illumination strategy.

\begin{table}[t]
\centering
\small
\caption{Comparison of lighting policies. Mean-PPFD denotes an iso-energetic baseline constructed by trace-driven simulation.}
\label{tab:lighting_sim}
\begin{tabular}{lccc}
\toprule
\textbf{Policy} & \textbf{$P_{n,\mathrm{int}}$ (mol m$^{-2}$)} & \textbf{Energy (kWh)} & \textbf{$\Delta(P_{n,\mathrm{int}}/E)$ (\%)} \\
\midrule
Mean-PPFD (Sim) & 7.106 & 8.255 & -11.0 \\
Rule-based & 9.396 & 13.824 & -30.2 \\
\textbf{LightFARM} & \textbf{8.039} & \textbf{8.255} & \textbf{--} \\
\bottomrule
\end{tabular}
\end{table}


\subsection{Photosynthesis Model Validation}
\label{subsec:pn_validation}

This subsection validates the photosynthesis predictor used in \cname{} and compares several lightweight regression models under the measured basil dataset. 
The goal is to justify the predictor adopted in the controller by evaluating prediction accuracy in different environmental conditions.

\paragraph{Data Collection and Measurement Protocol.}
Twenty healthy perennial basil plants were cultivated under controlled lighting, temperature, and irrigation conditions without fertilizer application. 
Leaf-level $P_\mathrm{n}$ was measured using a LI-6800 portable photosynthesis system to provide ground-truth labels~\cite{LICOR_LI6800}. 

Environmental variables were systematically varied across six temperature levels, ten PPFD levels, five R:B ratios, and two CO$_2$ concentrations, yielding 600 distinct measurement conditions. 
All other parameters were kept constant, including airflow and relative humidity. 
Table~\ref{tab:envsettings} summarizes the environmental settings. 
Measurements were taken on the third fully expanded leaf during daytime periods chosen to avoid strong diurnal artifacts, and only steady-state values were retained.

\begin{table}[t]
\centering
\caption{Environmental parameter settings for photosynthetic measurements.}
\label{tab:envsettings}
\begin{tabularx}{\linewidth}{lXc}
\toprule
\textbf{Factor} & \textbf{Gradient} & \textbf{Count} \\
\midrule
PPFD ($\mu$mol m$^{-2}$ s$^{-1}$) & 0, 50, 100, 200, 300, 400, 500, 600, 800, 1000 & 10 \\
R:B Ratio & 1:1, 3:1, 5:1, 7:1, R-only & 5 \\
CO$_2$ ($\mu$mol mol$^{-1}$) & 400, 800 & 2 \\
Temperature (°C) & 18, 20, 22, 24, 26, 30 & 6 \\
\bottomrule
\end{tabularx}
\end{table}

\paragraph{Model Comparison Results.}
We evaluated three lightweight regression models, including support vector regression (SVR), the MLP, and k-nearest neighbors (KNN), with Bayesian optimization (BO) or PSO used for hyperparameter tuning where applicable. 
Three-fold cross-validation was used for fair comparison.

Table~\ref{tab:pn_eval} summarizes the results. 
Among the evaluated configurations, the proposed MLP--PSO model achieved the best overall performance, with the lowest MSE, RMSE, and MAE, and the highest $R^2$. 
These results justify its use as the default photosynthesis predictor in the subsequent \cname{} experiments.

\begin{table}[t]
\centering
\caption{Comparison of photosynthesis prediction models under different optimization strategies.}
\label{tab:pn_eval}
\begin{tabular}{lccccc}
\toprule
\textbf{Model} & \textbf{Optimizer} & \textbf{MSE} & \textbf{RMSE} & \textbf{MAE} & $\boldsymbol{R^2}$ \\
\midrule
\textbf{MLP} & \textbf{PSO} & \textbf{0.5669} & \textbf{0.7530} & \textbf{0.5710} & \textbf{0.9859} \\
MLP & BO  & 0.6785 & 0.8237 & 0.6498 & 0.9832 \\
SVR & BO  & 0.7606 & 0.8721 & 0.6382 & 0.9811 \\
SVR & PSO & 1.0680 & 1.0335 & 0.7977 & 0.9735 \\
KNN & --   & 5.0713 & 2.2520 & 1.7496 & 0.8742 \\
\bottomrule
\end{tabular}
\end{table}




\subsection{Battery-Free Sensing Microbenchmark and Feasibility}
\label{subsec:batteryfree_evaluation}

This subsection evaluates the practical feasibility of the battery-free sensing subsystem under the deployed indoor lighting schedule. 
The goal is not to claim universal energy autonomy, but to determine whether the sensing workload required by \cname{} can be sustained from harvested optical energy in the tested chamber deployment.

\subsubsection{Energy Consumption}
\label{subsubsec:sensing_energy}

The node power was profiled using the Nordic Semiconductor Power Profiler Kit~II.
Its energy usage is dominated by sensing, wireless communication, and MCU activity, and the node alternates between two modes: \textit{Sleep Mode} and \textit{Active Mode}.

In \textit{Sleep Mode}, all sensors and communication interfaces are disabled and only the MCU real-time clock remains active.
This mode dominates the duty-cycled workload and consumes approximately \textbf{17.86~\textmu W}. 

In \textit{Active Mode}, the node performs one full sensing--transmission cycle, including spectral, temperature--humidity, and CO$_2$ measurements followed by BLE transmission.
This phase consumes approximately \textbf{4.38~mW}. 
Under the deployed configuration with a 30~s sensing interval and a 0.1~s active window, the node remains in \textit{Sleep Mode} for about 99.7\% of the time, resulting in an average power of roughly \textbf{33~\textmu W}.
Table~\ref{tab:node_current} summarizes the active-mode current draw of the main components. 
The results show that BLE transmission dominates the active-mode energy cost, while the average workload remains sufficiently low for harvested-energy operation under the tested lighting schedule.

\subsubsection{Harvesting and Storage Behavior}
\label{subsubsec:sensing_storage}

To assess long-term sensing feasibility in the deployed system, the node was operated with a 100~mF supercapacitor and its capacitor voltage $V_\mathrm{cap}$ was monitored over a 12-day cultivation run.
The node harvested optical energy directly from the LED illumination without wired power or battery assistance.

The measured $V_\mathrm{cap}$ exhibited a repeatable charge--discharge pattern synchronized with the illumination cycle. 
During light periods, the capacitor voltage increased as harvested energy accumulated; during dark periods, it decreased as the stored energy supported sensing and communication. 
Under the tested indoor lighting schedule and sensing workload, the node remained operational throughout the experiment, indicating that the harvested optical energy was sufficient to sustain the deployed sensing workload over the full 12-day run. These results therefore demonstrate \emph{sensing feasibility under the tested indoor lighting schedule}, rather than operation near a strict energy-neutral boundary.
In other words, the deployed battery-free node can reliably support the sensing workload required by \cname{} in the chamber-scale setup studied here.

\begin{table}[t]
\centering
\caption{Active-mode current consumption of the battery-free sensing node.}
\label{tab:node_current}
\begin{tabular}{lc}
\toprule
\textbf{Component} & \textbf{Current (\textmu A)} \\
\midrule
MCU (Idle) & 8.95 \\
CO$_2$ Sensor & 95.81 \\
Spectral Sensor & 64.93 \\
Temperature--Humidity Sensor & 9.60 \\
BLE Transmission & 1,530 \\
\bottomrule
\end{tabular}

\end{table}


\section{Discussion and Limitations}
\label{sec:discussion}

\subsection{Discussion}
\label{subsec:discussion_main}
Our results show that \cname{} moves indoor lighting toward a more energy-efficient operating region through predictive control. By jointly accounting for plant response, lighting energy use, and thermal effects, the controller reduces unnecessary illumination and improves energy productivity relative to rule-based control, albeit with an explicit energy--yield trade-off.

The results also show that battery-free sensing can be supported under the tested indoor lighting conditions without separate wired power. In the current design, sensing remains a secondary feasibility requirement, so the controller applies only a minimal correction when additional harvested energy is needed.

\subsection{Limitations and Future Work}
\label{subsec:limitations}
Several limitations remain.
First, the photosynthesis prediction model was trained only on perennial basil under a limited set of environmental and spectral conditions.
Extending it to other crops is challenging because photosynthetic response depends on species, growth stage, and plant-to-plant variability.
Perennial basil is also a relatively low-light-demand crop. The present results therefore should not be interpreted as a universal savings bound across species; future work should retrain and re-evaluate the same MPC+MLP framework on higher-light-demand crops such as lettuce or tomato.
In addition, reliable $P_\mathrm{n}$ data collection requires controlled gas-exchange measurements, which are time-consuming to obtain across multiple crops and developmental stages.
Second, the controller uses predicted $P_\mathrm{n}$ as a short-term proxy for productivity rather than directly optimizing long-term biomass or final yield.
Third, the LED thermal model is a simplified lumped representation that does not capture complex effects such as airflow heterogeneity, spatial heat transfer, or active HVAC interaction in larger deployments.
Finally, sensing feasibility is currently handled through a minimal correction layer rather than a fully unified joint optimal-control formulation.
Visual comfort is a secondary concern in the largely automated chamber setting studied here, but mixed-use facilities could employ a human-entry mode that temporarily freezes rapid dimming whenever personnel are present.

Future work will extend \cname{} to more diverse crops and larger indoor systems, improve model generalization, and develop more tightly integrated yield-aware and sensing-aware predictive control strategies.

\section{Related Work}
\label{sec:related_work}  

\subsection{Energy Challenges in Indoor Farming}
\label{subsec:energy_challenges}
Lighting remains the dominant source of energy consumption in indoor and vertical farming systems, typically accounting for over 60\% of total energy use~\cite{dauchot2024energy}.  
Most existing installations rely on fixed lighting schedules or empirically tuned rules that ignore dynamic plant responses and environmental variability~\cite{lempiainen2022plants}.  
Such rule-based control often leads to excessive energy expenditure or suboptimal growth performance.  
These limitations have motivated growing interest in adaptive, data-driven, and predictive control frameworks that can improve energy efficiency while maintaining crop productivity.

\subsection{Lighting Control Strategies}
\label{subsec:lighting_control}
Early studies on lighting control primarily focused on fixed-time schedules~\cite{pennisi2019unraveling} or rule-based feedback systems~\cite{serale2021supervisory} for regulating LED output. These approaches relied on empirical thresholds or environmental triggers (e.g., light sensors) to achieve basic automation, but lacked predictive capability and systematic optimization.

More recent work has explored heuristic optimization strategies~\cite{chen2023multi} to balance energy consumption and crop productivity, especially in systems without comprehensive dynamical models. Spectral tuning efforts~\cite{wang2022red} have further advanced LED system design by identifying optimal red–blue or multiband combinations under static control settings.

In contrast to these strategies, \cname{} integrates predictive plant-response modeling and thermal-aware actuation into a dynamic MPC framework, enabling real-time adaptive lighting control while jointly provisioning power to battery-free sensor nodes.


\subsection{Battery-Free Sensing and Sustainable IoT Systems}
\label{subsec:batteryfree_related}
To minimize maintenance costs and improve sustainability, recent work has explored energy-harvesting and battery-free sensor networks for smart agriculture and indoor environments. Systems such as SmartON~\cite{luo2021smarton}, PreAct~\cite{geissdoerfer2019getting}, and Riotee~\cite{geissdoerfer2024riotee} demonstrated the feasibility of intermittently powered sensing and programmable energy-aware operation in low-energy settings. Related work has also investigated low-power communication architectures relevant to indoor farming. For example, Javed et al.~\cite{javed2021rffree} use LED grow lights to support RF-free smart-farming communication through optical wireless links, while ambient RF backscatter systems~\cite{liu2013ambient} show how battery-free devices can reduce communication-side radio power. The patent in~\cite{li2022cea} similarly considers adapting LED intensity for plant and environmental monitoring. These efforts are complementary to our setting, but they primarily address communication architecture or monitoring support rather than integration with a plant-aware predictive control loop. In contrast, \cname{} studies how a photovoltaic-powered battery-free sensing node with supercapacitor storage and BLE-based communication can be incorporated into an indoor lighting control framework driven by photosynthesis and thermal models. Rather than claiming the first battery-free IoT system for indoor farming, our contribution lies in demonstrating a chamber-scale integration of harvested-energy sensing with predictive lighting control, while explicitly evaluating sensing feasibility under the tested indoor lighting schedule.

\section{Conclusion}
\label{sec:conclusion}

This paper presented \cname{}, an integrated cyber-physical system that combines MPC with battery-free sensing for energy-efficient lighting management in indoor farming.  
By leveraging a data-driven photosynthesis prediction model, a lightweight thermal model, and a sensing-energy feasibility model, \cname{} dynamically regulates LED intensity while accounting for plant response, energy use, and thermal constraints.  
In the proposed co-design, plant-oriented predictive control remains primary, and sensing sustainability is enforced through a minimal feasibility correction when needed. Experimental results on a physical indoor-farming testbed across two independent 12-day basil runs demonstrated that \cname{} reduces lighting energy by approximately 40--42\% and improves energy productivity by about 46\% relative to the rule-based baseline.
These gains were accompanied by reduced fresh biomass, indicating an explicit energy--yield trade-off under the tested conditions.  
Overall, the results support \cname{} as a practical approach to improving indoor-farming energy efficiency while maintaining battery-free sensing under shared LED infrastructure.


\bibliographystyle{ACM-Reference-Format}
\bibliography{reference}

\end{document}